# *Numerical study of The Remittances of Axially Excited Chiral Sculptured Zirconia Thin Films*

Ferydon Babaei and Hadi Savaloni[*]

*Department of Physics, University of Tehran, North-Kargar Street, Tehran, Iran.*
*\*) Corresponding author: Tel: +98 21 6635776; Fax: +98 21 88004781; Email: savaloni@khayam.ut.ac.ir*

**Abstract**

The transmission and reflection spectra from a right-handed chiral sculptured zirconia thin film are calculated using the coupled wave theory and the Bruggeman homogenization formalism in conjunction with the experimental data for the relative dielectric constant of zirconia thin film. The dielectric dispersion function effect on these spectra appeared in wavelengths shorter than the Bragg wavelength. In wavelengths larger than the Bragg wavelength, the dispersion of the dielectric function can be ignored. The results achieved in this work are consistent with the experimental data (Wu *et al.* (2000)). A shift towards shorter wavelengths is observed for the Bragg peak with increasing the void fraction, which is in agreement with the theoretical work of Lakhtakia (2000). Sorge *et al.* (2006) also found this effect in their experimental results on $TiO_2$ chiral thin films, while they also found that unlike our results the intensity of the reflectance of the Bragg peak decreases with increasing the void fraction. This difference between our theoretical work and Sorge *et al.*'s (2006) experimental work can be related to the structural difference between idealized theoretical model for chiral films and that obtained in experimental work. In the latter, as Sorge *et al.* (2006) have pointed out the experimental films exhibit a large amount of scattering due to the highly complex and non-ideal structure that the individual chiral elements exhibit.

# 1. Introduction

In obtaining the transmission and reflection spectra for sculptured thin films, the relative dielectric constant of the film material at a certain frequency is usually considered, and then Bruggeman homogenization formalism is used to estimate the relative permittivity scalars. These scalar quantities are assumed constant in the procedure of obtaining the transmission and reflection spectra at all frequencies [1-3]. However, it is clear that the relative dielectric constant is dependent on the frequency and at each frequency it has a particular value. In addition, in most of the reports on the dielectric dispersion function effect on remittances (reflection and transmission) of sculptured thin films, the simple single-resonance Lorentzian model is used [4-6], but when this model is used there exist some difficulties in obtaining the oscillator strengths, resonance wavelengths, and absorption line widths,.

The aim of this work is to report on the real dispersion effect in chiral sculptured zirconia thin film, by using the experimental data of relative dielectric constant of thin film at each frequency [7] in conjunction with the coupled wave theory and the Bruggeman homogenization formalism and to show that this effect is pronounced at wavelengths smaller than the Bragg wavelength, consistent with the experimental results [8]. It should be mentioned here that the refractive index for zirconia in the wavelength region examined in this work has only real value and the imaginary part is zero [7]. Since there has been no experimental data on zirconia we have been forced to compare our results with those of titanium oxide thin films. Currently, we are working on films with complex refractive index, so that the results can be directly compared with the available experimental results such as titanium oxide films. The results of these two separate works will no doubt provide us with the pure influence of the imaginary part of the refractive index in this kind of study.

## 2. Theory

Assume that the space ($0 \leq z \leq d$) is occupied by a right-handed chiral sculptured thin film and $z \leq 0$ and $z \geq d$ have refraction indices of $n_{inc}$ and $n_{trans}$, respectively (Fig. 1) [9]. The application of the coupled wave theory allows us to represent the complete field transfer between incident media and the transfer media for axial propagation of plane waves in the direction of inhomogenity axis of the thin film (z-axis) in normal incidence in a Matrix form [10-12]:

$$\varepsilon\big|_{z=d^+} = N(n_{av}, n_{trans}).S(n_{av}, \delta n, \kappa_0, \Omega, d).N(n_{inc}, n_{av}).\varepsilon\big|_{z=0^-} \qquad (1)$$

where:

$$N(\upsilon_a, \upsilon_b) = \frac{1}{2}\begin{bmatrix} 1+\frac{\upsilon_a}{\upsilon_b} & 0 & 0 & 1-\frac{\upsilon_a}{\upsilon_b} \\ 0 & 1+\frac{\upsilon_a}{\upsilon_b} & 1-\frac{\upsilon_a}{\upsilon_b} & 0 \\ 0 & 1-\frac{\upsilon_a}{\upsilon_b} & 1+\frac{\upsilon_a}{\upsilon_b} & 0 \\ 1-\frac{\upsilon_a}{\upsilon_b} & 0 & 0 & 1+\frac{\upsilon_a}{\upsilon_b} \end{bmatrix} \qquad (2)$$

$$S = \begin{bmatrix} e^{i\kappa d} & 0 & 0 & 0 \\ 0 & P & 0 & Q \\ 0 & 0 & e^{-i\kappa d} & 0 \\ 0 & Q^* & 0 & P^* \end{bmatrix} \qquad (3)$$

the asterisk denotes the complex conjugate, and:

$$\begin{cases} P = e^{ipd}[\cosh(\Delta d) + i\frac{\delta\kappa}{2\Delta}\sinh(\Delta d)], \\ Q = \frac{ik}{\Delta}e^{ipd}\sinh(\Delta d), \\ \Delta = +[|k|^2 - (\frac{\delta\kappa}{2})^2]^{1/2}. \end{cases} \qquad (4)$$

$$\varepsilon = \begin{bmatrix} E_L^+ \\ E_R^+ \\ E_L^- \\ E_R^- \end{bmatrix} \qquad (5)$$

where $p = \dfrac{\pi}{\Omega}$ and $\Omega$ is the half of the structural period. In axial propagation two refraction indices become important [12]:

$$n_c = \varepsilon_c^{1/2} \quad , \quad \tilde{n}_d = \tilde{\varepsilon}_d^{1/2} = \left(\dfrac{\varepsilon_a \varepsilon_b}{\varepsilon_a \cos^2 \chi + \varepsilon_b \sin^2 \chi}\right)^{1/2} \qquad (6)$$

$\varepsilon_{a,b,c}$ are the relative permittivity scalars, $\tilde{\varepsilon}_d$ is the composite relative permittivity scalar and $\chi$ is the angle of rise of the chiral nanostructure of which the film is comprised [9]. The refractive index of the chiral sculptured thin film is given by [12]:

$$n_{av} = \left(\dfrac{\tilde{n}_d + n_c}{2}\right)$$

and the birefringence in this CSTF is introduced as:

$$\delta n = (\tilde{n}_d - n_c)$$

In these expressions $k = \pi \delta n / \lambda_0$, $\delta\kappa = 2(\kappa - p)$, and $\kappa = 2\pi n_{av} / \lambda_0$ and $\lambda_0$ is the free space wavelength.

The solution of the Matrix equation (1) with $E_L^- \big|_{z=d^+} = E_R^- \big|_{z=d^+} = 0$ condition provides us with the electrical fields for reflection and transmission and we can obtain the reflection and transmission coefficients as:

$$\begin{cases} r_{i,j} = \dfrac{E_i^- \big|_{z=0^-}}{E_j^+ \big|_{z=0^-}} \\ t_{i,j} = \dfrac{E_i^+ \big|_{z=d^+}}{E_j^+ \big|_{z=0^-}} \end{cases} , i,j = L, R \qquad (7)$$

the reflectance and transmittance can be calculated as:

$$\begin{cases} R_{i,j} = |r_{i,j}|^2 \\ T_{i,j} = |t_{i,j}|^2 \end{cases}, i, j = L, R \tag{8}$$

straightforward relations for reflectance and transmittance coefficients can be obtained from the coupled wave theory, when $n_{inc} = n_{trans} = n$:

$$\begin{cases} r_{LL} = \dfrac{2ikn n_{av}(n_{av}-n)^2}{D\Delta} e^{i(\kappa+p)d} \sinh(\Delta d), \\ r_{LR} = r_{RL} = \dfrac{n_{av}^2 - n^2}{D}\{(n_{av}^2 + n^2)[1 - \mathrm{Re}(e^{i\kappa d}P)] + 2i\,n n_{av}\,\mathrm{Im}(e^{i\kappa d}P)\}, \\ r_{RR} = \dfrac{2ikn n_{av}(n_{av}+n)^2}{D\Delta} e^{-i(\kappa+p)d} \sinh(\Delta d). \end{cases} \tag{9}$$

$$\begin{cases} t_{LL} = \dfrac{2n n_{av}}{D}[(n_{av}+n)^2 P^* - (n_{av}-n)^2 e^{i\kappa d}], \\ t_{LR} = t_{RL} = \dfrac{2\,ikn n_{av}(n_{av}^2 - n^2)}{D\Delta} e^{-i\kappa d} \sinh(\Delta d), \\ t_{RR} = \dfrac{2n n_{av}}{D}[(n_{av}+n)^2 e^{-i\kappa d} - (n_{av}-n)^2 P]. \end{cases} \tag{10}$$

where Re( ) and Im( ) are the real and imaginary parts of the quantity given in the parenthesis, respectively.

A medium is generally dispersive and the propagation of waves in a medium is influenced by dispersion [13]. In order to investigate the dispersion effect on remittances (reflection and transmission) of sculptured thin films, usually the single-resonance Lorentzian model is implemented [4-6]. However, in this work, in order to investigate the effect of dispersion on remittances of sculptured thin films, we have used the experimental data of dielectric refractive index for zirconia [7] and the fact that the refraction coefficient of the material used $n_s(\omega) + ik(\omega)$, in its bulk form depends on the frequency, and the relative dielectric constant is $\varepsilon_s(\omega) = (n_s(\omega) + ik(\omega))^2$.

In order to obtain the relative permittivity scalar $\varepsilon_{a,b,c}$ in sculptured thin films in each frequency, the Bruggeman homogenization formalism was used [1]. In this formalism the film is considered as a two phase composite, namely void phase and the material (inclusion) phase. Both of these are dependent on column form factor, the fraction of void phase, free space wavelength and the refractive index of the material. In addition each column in the STF structure is considered as a string of identical long ellipsoids (Fig. 2) [9]. The ellipsoids are considered to be electrically small (i.e., small in a sense that their electrical interaction can be ignored) [1-3]. Therefore:

$$\varepsilon_\sigma = \varepsilon_\sigma(\varepsilon_s(\lambda_0), f_v, \gamma_\tau^s, \gamma_b^s, \gamma_\tau^v, \gamma_b^v), \qquad \sigma = a,b,c,d \qquad (11)$$

where $f_v$ is the fraction of void phase, $\gamma_\tau^{s,v}$ is one half of the long axis of the inclusion and void ellipsoids, and $\gamma_b^{s,v}$ is one half of the small axis of the inclusion and void ellipsoids.

### 3. Numerical results and discussion

It was assumed that a right-handed zirconia chiral sculptured thin film in its bulk state is formed, which occupies the space in the free space ($n=1$) with a thickness of $d$. In order to obtain the relative permittivity scalars we used the Bruggeman homogenization formalism with columnar form factors $\gamma_\tau^s = 20$, $\gamma_b^s = 2$, $\gamma_\tau^v = 1$, $\gamma_b^v = 1$. Setting the shape factors $\gamma_\tau^s \gg 1$ and $\gamma_b^s \gg 1$ will make each ellipsoid resemble a needle with a slight bulge in its middle part [9]. With regard to voids these needles are assumed to take up spherical shapes.

A range of wavelengths $\lambda_0 \in (250nm - 850nm)$ was considered, where the real refractive index of zirconia in its bulk state (Fig. 3) varies from 2.64599 to 2.17282 for the lowest wavelength to highest wavelength, respectively [7]. It should be noted that the imaginary part of the refractive index for zirconia in this range of the wavelengths is zero (Fig. 3). The relative permittivity scalar at each frequency was obtained using the Bruggeman homogenization formalism. It is obvious that by knowing these scalars and the rise angle $\chi$, one can calculate the two refractive indices $n_c$ and $\tilde{n}_d$ using Eq. (6). Application of Eqs. (9 and 10) will provide us with the reflectance and transmittance coefficients.

In Fig. 4 the reflectance and transmittance from a right-handed zirconia CSTF with following parameters: $d = 40\Omega$, $\Omega = 150\,nm$, $\chi = 30^0$ for different void fractions are depicted. From the $R_{LL}$ and $R_{RR}$ plots in Fig. 4, it can be deduced that the circular Bragg phenomenon vanishes at $f_v = 0$ and $f_v = 1$, because CSTF becomes isotropic ($\varepsilon_a = \varepsilon_b = \varepsilon_c$) and homogenized. For void fractions values between $f_v = 0.2$ and $f_v = 0.8$, the Bragg wavelength assumes values which are given in Table 1. The resultant reflectances are also presented in Table 1 and Fig.4 ($R_{RR}$ plots).

Table 1. Void fractions and the corresponding values obtained for Bragg wavelength and the spectral reflectance $R_{RR}$ zirconia CSTF

| $f_v$ | 0.2 | 0.4 | 0.6 | 0.8 |
|---|---|---|---|---|
| $\lambda_o^{Br}$ | 590 | 520 | 440 | 370 |
| $R_{RR}$ | 0.050 | 0.308 | 0.880 | 0.973 |

This shows that as the void fraction increases the Bragg wavelength shifts towards shorter wavelengths. In addition the reflectance increases, for by increasing the void phase, the interference becomes constructive. Our result with respect to the shift of the Bragg wavelength towards shorter wavelengths by increasing the void fraction is in agreement with Sorge *et al.*'s experimental results on TiO$_2$ chiral films [14]. However, there exist a difference between our theoretical (numerical) work,

Lakhtakia's theoretical work and that of Sorge *et al.*'s experimental work: the reflectance in $R_{RR}$ increases with void fraction in our work and that of Lakhtakia [3], while it is decreased with increasing the void fraction in Sorge *et al.*'s work [14]. Robbie *et al.* [15] reported the impossibility of fabrication of real chiral thin films with high void fraction without depositing the films at a high deposition angle. Therefore, there exist a pronounced structural difference between the idealized theoretical model for chiral thin films and that obtained in the experimental work. The experimental thin films consist of highly complex and non-ideal structures, which can be the source of the large amount of scattering reported by Sorge *et al.* [14]. This scattering effect is the cause of the discrepancy between our theoretical work and Sorge *et al.*'s experimental results.

The reason for the small values of reflectance in $R_{RR}$ plots is that the handedness of thin film is not the same as the polarization of the light.

In $T_{LL}$ and $T_{RR}$ plots, the transmittance in the Bragg region is small, while it increases out side this region. The contributions of the short wavelength region of the measured co-polarization transmission spectra ($T_{LL}$ and $T_{RR}$ plots) in a CSTF are related to the dispersion [8]. In our results, this effect occurs at wavelengths region smaller than Bragg wavelength. Therefore, it may be concluded that for wavelengths larger than Bragg wavelength, the dispersion of dielectric function is negligible and its value can be considered as constant.

In the plot of $R_{LR}$ it can be seen that the reflectance spectra decreases by increasing the void fraction, and for $f_v = 1$, as expected, it becomes zero (see plot of $T_{RR}$ in Fig. 4). It can also be observed that the spectral dispersion increases at shorter wavelengths.

The maximum of $T_{LR}$ occurs at medium values of void fraction ($\sim f_v = 0.6$), and decreases at both sides towards the minimum ($f_v = 0$) and maximum ($f_v = 1$) of void fraction. This is due to the isotropic structural property of these films.

In Fig. 5 the reflectance and transmittance from a right-handed zirconia CSTF with following parameters: $f_v = 0.6$, $\chi = 30°$, $\Omega = 150 nm$ and for different thicknesses are depicted. The $R_{LL}$ and $R_{RR}$ results clearly show that the Bragg wavelength remains constant (440 nm) with increasing film thickness, but the circular Bragg phenomenon increases (intensifies) considerably, because the number of reflecting planes increases with thickness. This increase is greater for a layer whose structural handedness is the same as the polarization of the light. The transmission plots of co-polarized light ($T_{LL}$ and $T_{RR}$) show that in the wavelength region smaller than Bragg wavelength the dispersion occurs as disordered ripples, but at higher values than Bragg wavelength they have an oscillatory behavior about a constant value. Therefore, again it can be concluded that for wavelengths larger than Bragg wavelength, the dispersion of dielectric function in a CSTF is negligible.

In $R_{LR}$ plot, at longer wavelengths with increasing film thickness the consecutive peaks become narrower, hence their frequency increases. In the $T_{LR}$ plot the position of transmittance peak is fixed but its value increases with film thickness.

The transmittance and reflectance of a right-handed zirconia CSTF with $f_v = 0.6$, $d = 40\Omega$, $\Omega = 150 nm$ for different rise angles, $\chi$ are given in Fig. 6. In $R_{LL}$ and $R_{RR}$ plots, it is obvious that the value of reflectance and the position (wavelength) for Bragg peaks remains almost the same when the rise angle is increased from 10º to 30º. Therefore, these results show that the rise angle has no significant effect on the dispersion in CSTFs. However, its smaller values are preferable, as it is pointed out by Lakhtakia [16] in implementation of stand-alone optical pressure sensors. The dispersion effect can be clearly observed in co-polarized transmittance spectra ($T_{LL}$ and $T_{RR}$) as short wavelength contributions in the wavelength region smaller than the Bragg wavelength.

In the $R_{LR}$ plot, the spectral dispersion at shorter wavelength can be clearly observed. In the $T_{LR}$ plot the position and the intensity of the transmittance's peak remains almost constant, but its width decreases with increasing the rise angle.

In Fig. 7 the transmittance and reflectance from a right-handed zirconia CSTF with $f_v = 0.6$, $d = 40\Omega$, $\chi = 30°$ for different values of half structural period, $\Omega$ are presented. The position (wavelength) of Bragg peak shifts to longer wavelengths with increasing the half structural period, because $\lambda_\circ^{Br} \approx 2\Omega n_{av}$ and the fact that in order to have constructive interference, light must travel at least one structural period. With reference to the $T_{LL}$ and $T_{RR}$ plots it can be deduced that for wavelengths longer than Bragg wavelength, the dispersion of dielectric function is insignificant and the dielectric constant of a CSTF at a given frequency at these wavelengths is constant, hence other properties of the CSTF can be calculated.

Again, the spectral dispersion at shorter wavelengths is obvious for the $R_{LR}$ data in this figure. In the $T_{LR}$ plot, by increasing the structural half period of the film the intensity of transmittance peak reduces and its width become broader, while its position shifts towards longer wavelengths.

In summary, in this work we have been able to show that the dispersion effect is more pronounced at wavelengths smaller than the Bragg wavelength (as short wavelength contributions in co-polarized transmittance spectra). This is consistent with the experimental results presented by Wu et al. [8], though their results are for structurally left-handed CSTF made of titanium oxide. This was performed by using the experimental refractive index of zirconia thin film [7] in conjunction with the coupled wave theory and the Bruggeman homogenization formalism at each given frequency. We should emphasize here that we did not use the simple single-resonance Lorantzian model which is usually used in order to describe the dispersion effect of relative dielectric permittivity scalar in a CSTF. This is because to our opinion it is possible that all layers of the film are not supposed to obey from single-resonance Lorentzian model, but they may be subject to double-resonance [17] or multi-resonance [18,19].

## 4. Conclusions

Using the coupled wave theory and the Bruggeman homogenization formalism the transmittance and reflectance of a right-handed zirconia CSTF were computed. This was carried out by considering the refractive index of zirconia (with zero value for the imaginary part in the frequency range examined in this work) at each given frequency, individually, in the frequency range of 250 to 850 nm in the homogenization formalism. Therefore in this way dispersion of the dielectric function was introduced into our calculations. This method takes advantage from the experimental relative dielectric constant of thin film and avoids the use of simple single-resonance Lorentzian model. In addition the reflectance and transmittance spectra are more realistic in a sense that the dispersion effect in computed co-polarized transmittance spectra occur in the same wavelength region (i.e., short wavelengths) as that in experimental results. Through out this work, as it is expected, it was possible to deduce that the contribution of dispersion is dominant in the wavelengths region smaller than Bragg wavelength and for wavelengths greater than Bragg wavelength one can ignore the dispersion effect and in this wavelength region one can assume that the dielectric function acts as a single frequency (constant). The results showed that the Bragg peak shifts towards shorter wavelengths with increasing the void fraction. This is consistent with the theoretical work of Lakhtakia [3], and the experimental work of Sorge *et al.* [14]. However, Sorge *et al.* [14] found a decrease in the intensity of the reflectance in Bragg peak with increasing the void fraction, while our results showed that the reflectance increases with void fraction. This is due to the scattering by the non-ideal and highly complex structure of experimentally grown chiral thin films.

**Acknowledgements**

This work was carried out with the support of the University of Tehran and the Iran National Science Foundation (INSF).

**Figure captions**

Figure 1. Schematic of the structural period $2\Omega$ and the angle of rise $\chi$ for a right-handed (RH) CSTF.

Figure 2. A STF column which is considered as a string of identical long electrical ellipsoids with shape factors of $\gamma_b$ and $\gamma_\tau$.

Figure 3. The refractive index of pure bulk zirconia, showing that the imaginary part is zero in the wavelength region shown.

Figure 4. Reflectance and transmittance from a right-handed zirconia CSTF with different void fractions. a) Reflectance, b) Tranmittance. $d = 40\Omega$, $\Omega = 150\,nm$, $\chi = 30^0$

Figure 5. Reflectance and transmittance from a right-handed zirconia CSTF with different thicknesses. a) Reflectance, b) Tranmittance. $f_v = 0.6$, $\chi = 30°$, $\Omega = 150\,nm$

Figure 6. Reflectance and transmittance from a right-handed zirconia CSTF with different rise angles. a) Reflectance, b) Tranmittance. $f_v = 0.6$, $d = 40\Omega$, $\Omega = 150\,nm$

Figure 7. Reflectance and transmittance from a right-handed zirconia CSTF with different half structural periods. a) Reflectance, b) Tranmittance. $f_v = 0.6$, $d = 40\Omega$, $\chi = 30°$

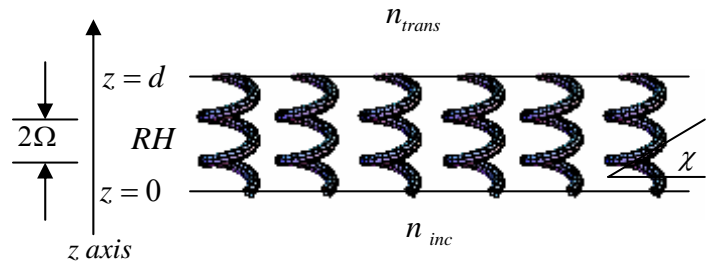

**Fig. 1; F. Babaei and H. Savaloni**

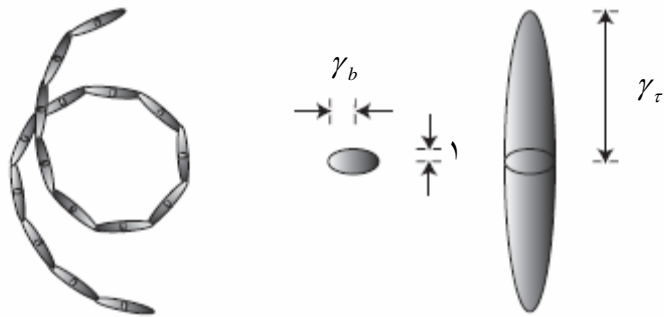

**Fig. 2; F. Babaei and H. Savaloni**

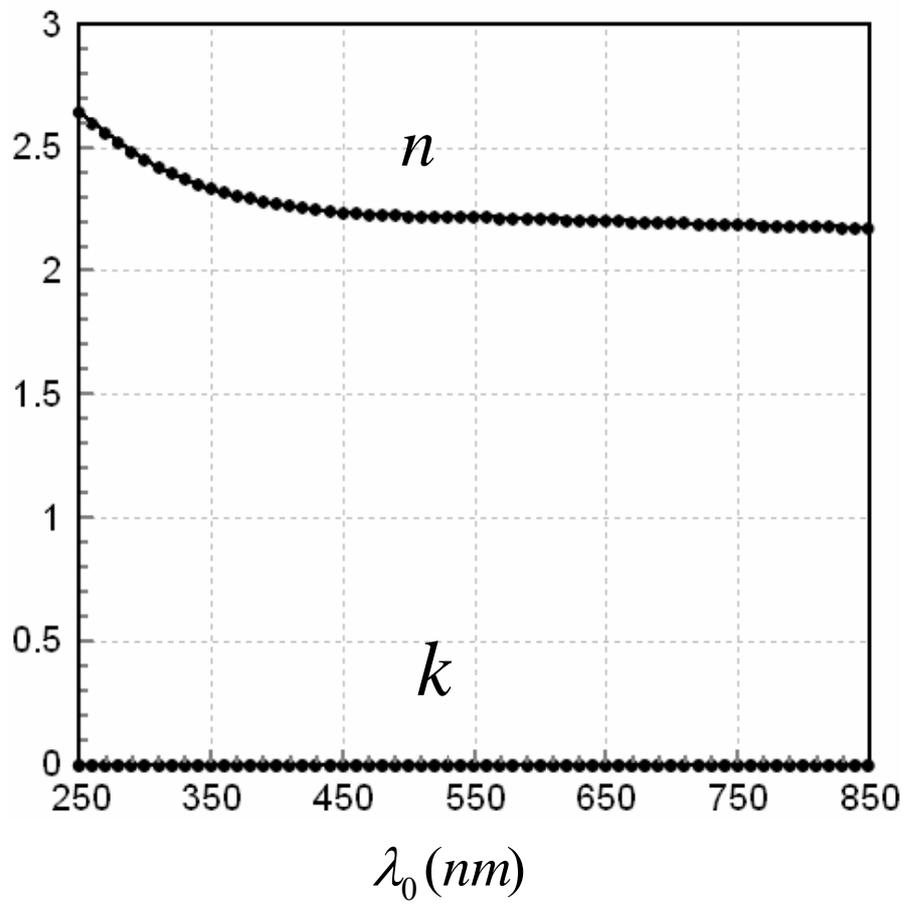

**Fig. 3; F. Babaei and H. Savaloni**

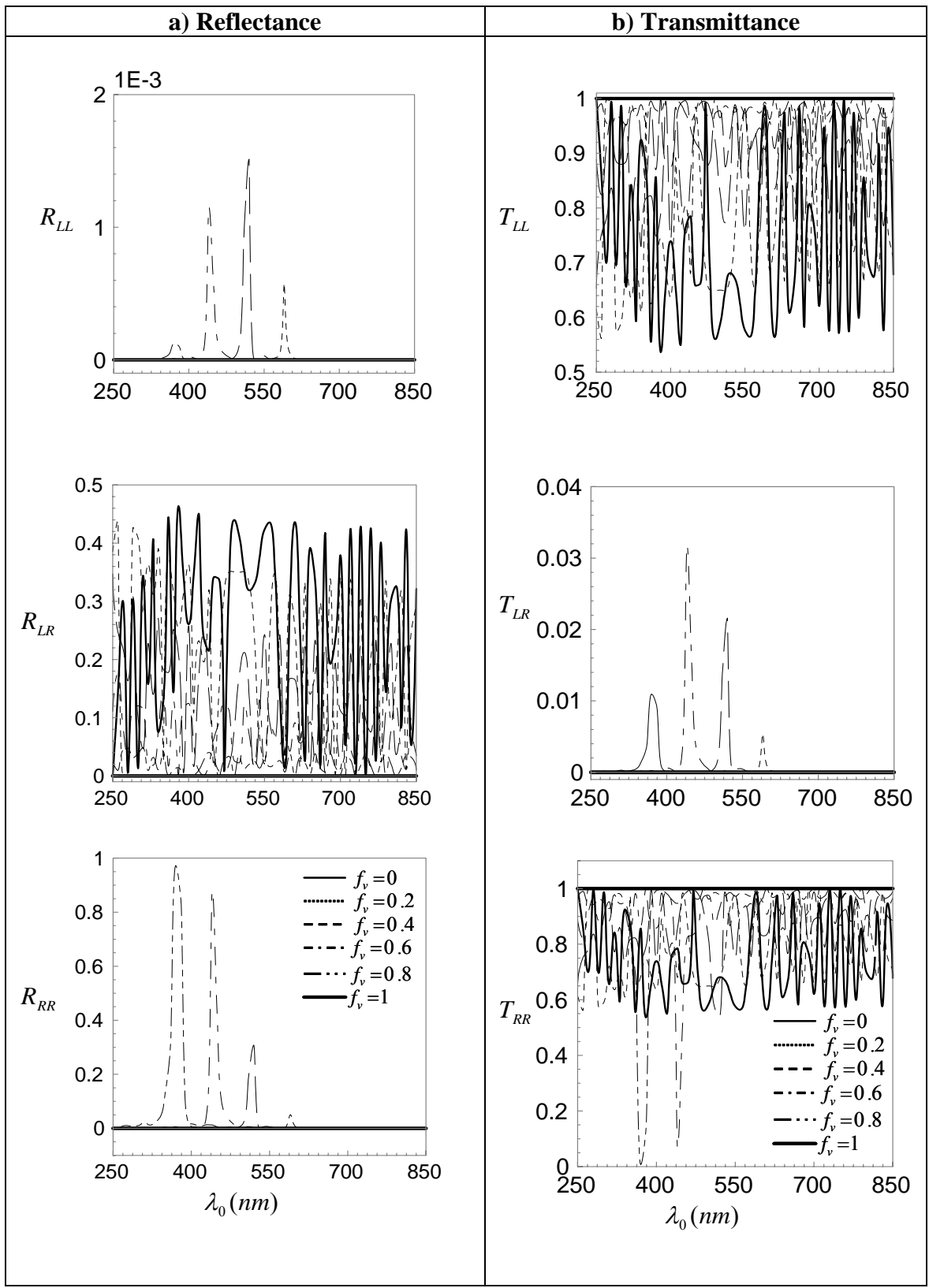

**Fig 4. F. Babaei and H. Savaloni**

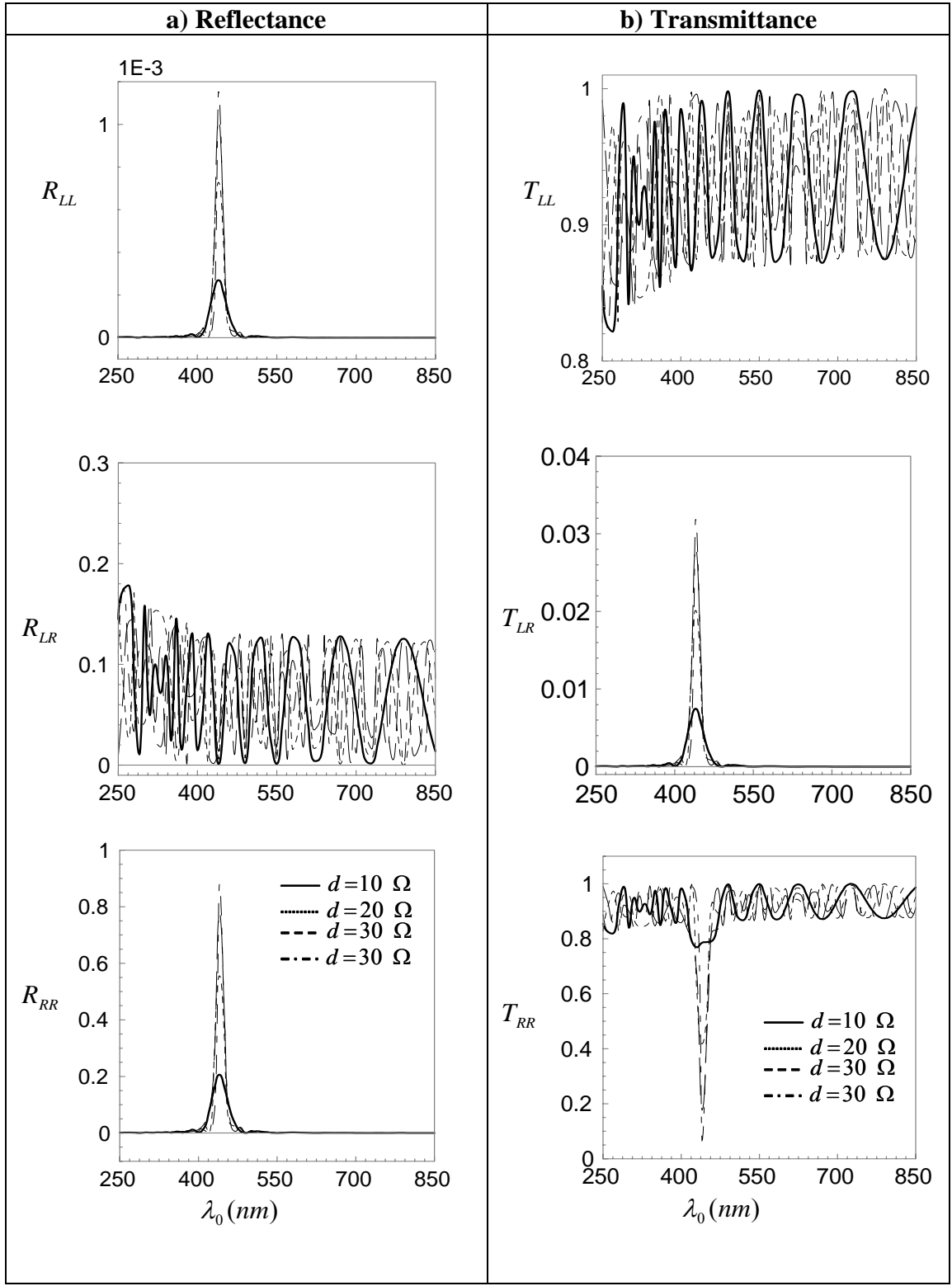

**Fig 5. F. Babaei and H. Savaloni**

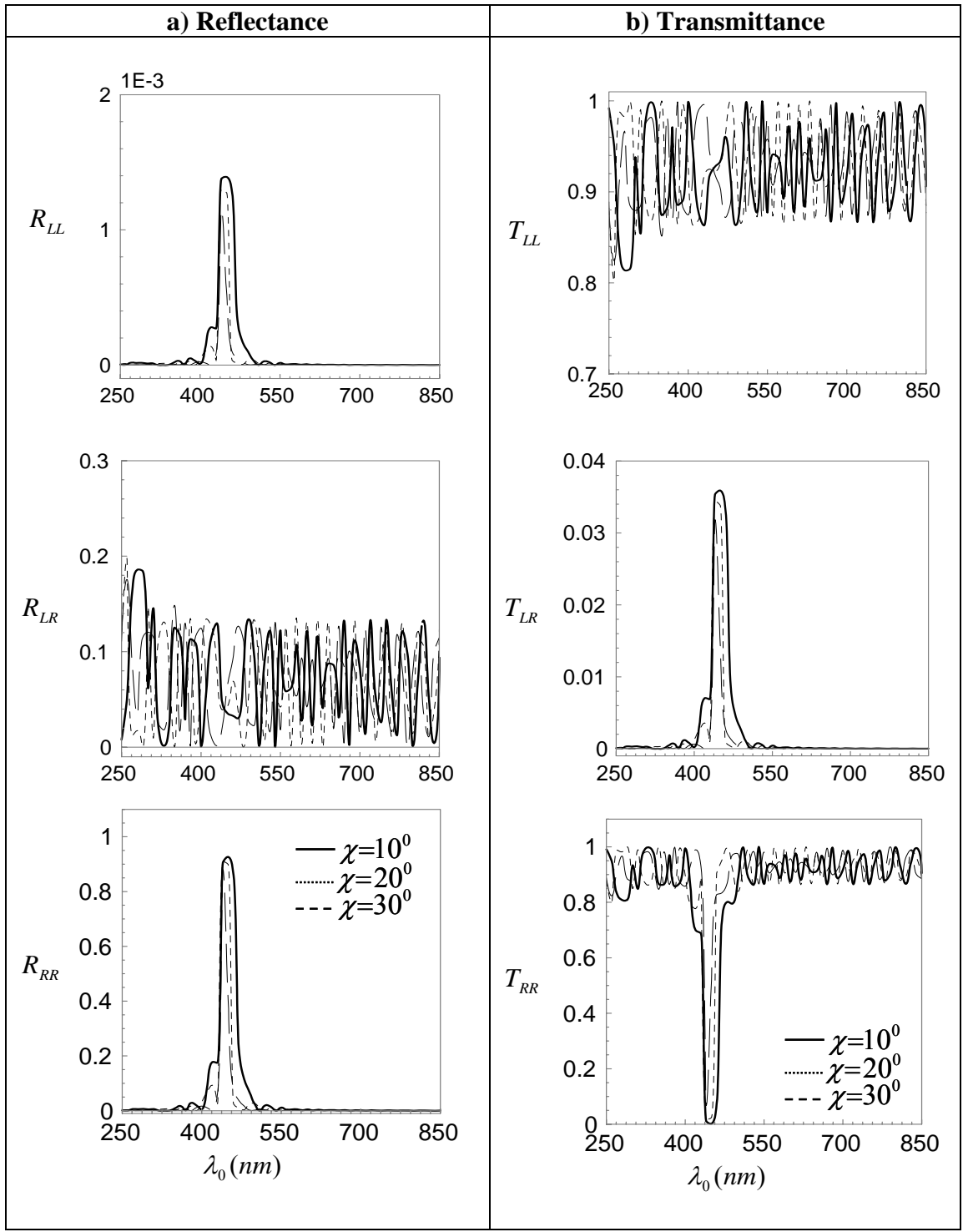

**Fig 6. F. Babaei and H. Savaloni**

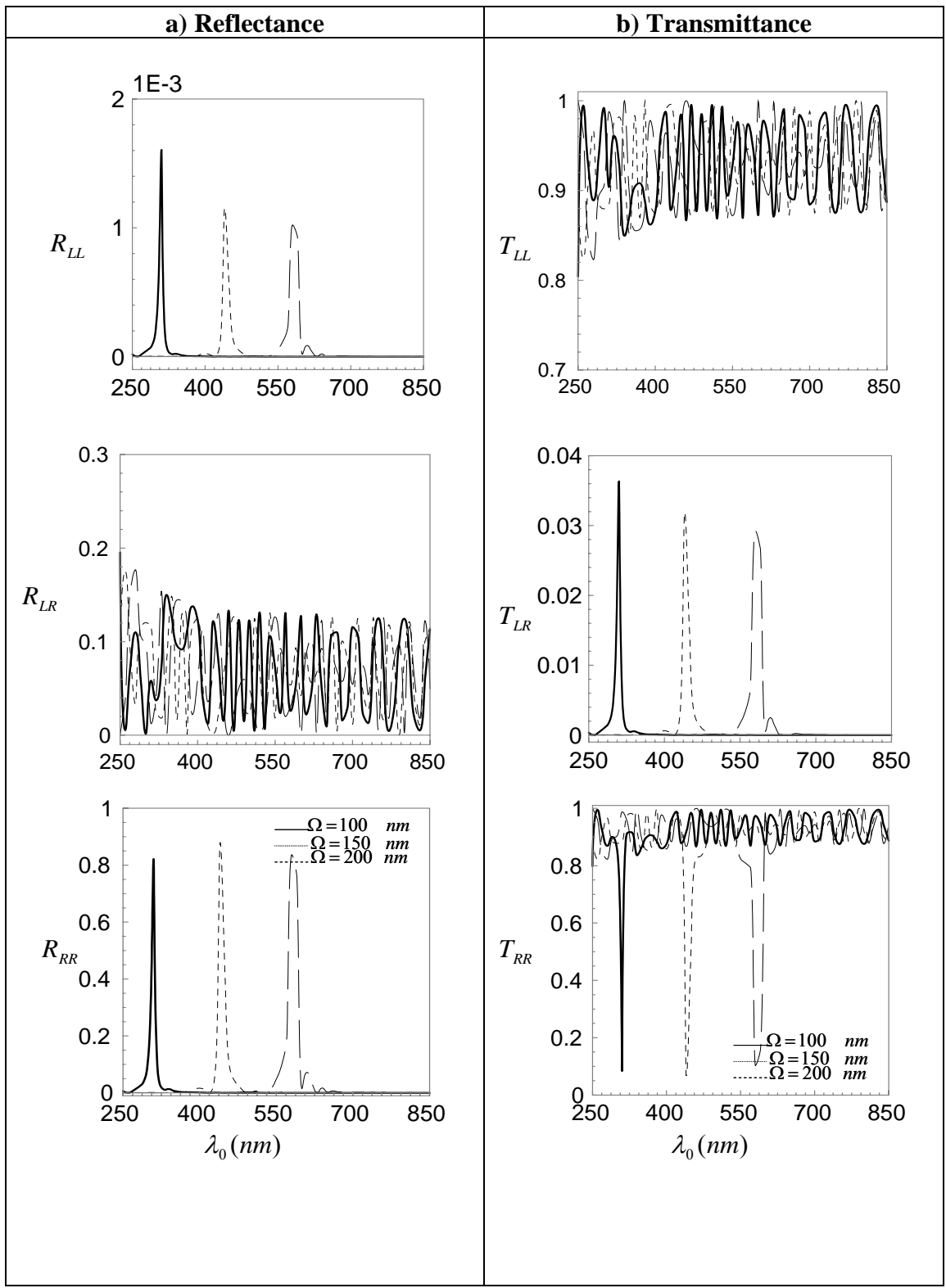

**Fig 7. F. Babaei and H. Savaloni**